\documentclass[
  numbers
    ,final            
  ]
  {aipproc}

\layoutstyle{8x11single}

\newcommand{\be}{\begin{equation}}
\newcommand{\ee}{\end{equation}}

\usepackage{multirow}
\usepackage{amsmath}
\usepackage{tabularx}
\usepackage{booktabs} 
\usepackage{caption} 

\begin{document}

\title{Regge trajectories of ordinary and non-ordinary mesons from their scattering poles}


\author{J.~Nebreda$^{*,\dagger,}$}{
  address={Yukawa Institute for Theoretical Physics, Kyoto University, 606-8502 Kyoto, Japan},
  altaddress={Center for Exploration of Energy and Matter, Indiana University, Bloomington, IN 47403, USA},
  address={Physics Department  Indiana University, Bloomington, IN 47405, USA },
  altaddress={Departamento de F\'isica Te\'orica II, Universidad Complutense de Madrid, 28040 Madrid, Spain}
}

\author{J.A.~Carrasco}{
  address={Departamento de F\'isica Te\'orica II, Universidad Complutense de Madrid, 28040 Madrid, Spain}
}

\author{J.T.~Londergan}{
  address={Center for Exploration of Energy and Matter, Indiana University, Bloomington, IN 47403, USA},
  altaddress={Physics Department  Indiana University, Bloomington, IN 47405, USA },
}

\author{J.R.~Pelaez}{
  address={Departamento de F\'isica Te\'orica II, Universidad Complutense de Madrid, 28040 Madrid, Spain}
}

\author{A.P.~Szczepaniak$^{\dagger,}$}{
  address={Center for Exploration of Energy and Matter, Indiana University, Bloomington, IN 47403, USA},
  address={Physics Department  Indiana University, Bloomington, IN 47405, USA },
 altaddress={Jefferson Laboratory, 12000 Jefferson Avenue, Newport News, VA 23606, USA }
}

\begin{abstract}
 Our results on obtaining the Regge trajectory of a resonance from its pole in a scattering process and from analytic constraints in the complex angular momentum plane are presented. The method, suited for resonances that dominate an elastic scattering amplitude, has been applied to the $\rho(770)$, $f_2(1270)$, $f_2«(1525)$ and $f_0(500)$ resonances. Whereas for the first three we obtain linear Regge trajectories, characteristic of ordinary quark-antiquark states, for the latter we find a non-linear trajectory with a much smaller slope at the resonance mass. We also show that if a linear trajectory with a slope of typical size is imposed for the $f_0(500)$, the corresponding amplitude is at odds with the data. This provides a strong indication of the non-ordinary nature of the sigma meson.
\end{abstract}

\maketitle


\section{Introduction}\label{aba:sec1}

We present here the results of a recent work~\cite{Londergan:2013dza} and an on-going project~\cite{Prep} were we use the analytic properties in the complex angular momentum plane to study the Regge trajectories of resonances that are predominantly elastic. 

It is a well-known experimental fact that when the angular momentum of the hadronic resonances is plotted versus their squared mass, almost all of them fall in linear trajectories with approximately the same slope. These trajectories can be intuitively understood in terms of rotational states of quarks linked by flux tubes. Thus the fact that some resonances, such as the $f_0(500)$, cannot be accommodated in any known Regge trajectory can be an indication that they have a non-ordinary nature. In this work, instead of trying to put the resonances into known trajectories, we develop a method to obtain the complex Regge trajectories of a predominantly-elastic resonance from the position and residue of its corresponding pole as it appears in the scattering of two other hadrons. This way we take into account the width of the resonance, in contrast with the previous works, which at most consider it as an error in the mass.

In~\cite{Londergan:2013dza} we applied this method to the lightest mesonic resonances appearing in $\pi\pi$ scattering, namely, the $\rho(770)$, which is well understood as a quark-antiquark state, and the $f_0(500)$ or $\sigma$ meson, whose nature is still the subject of a longstanding debate and which does not seem to fit well in the $(J,M^2)$ trajectories \cite{Anisovich:2000kxa}. Now, we are studying~\cite{Prep} the trajectories of two $J=2$ mesons, the $f_2(1270)$ and the $f_2'(1525)$, which decay predominantly into two pions and into two kaons, respectively.

\section{Regge trajectory of a resonance from its pole}

Near a Regge pole the partial wave can be written as
\be
t_l(s)  = \beta(s)/(l-\alpha(s)) + f(l,s),
\label{Reggeliket}
\ee
where $f(l,s)$ is a background regular function of $l$, and the Regge trajectory $\alpha(s)$ and 
residue $\beta(s)$ are analytic functions, the former having a cut along the real axis above the elastic threshold. If the pole dominates the scattering amplitude in the resonance region, the unitarity condition implies that, for real $l$, 
\be
\mbox{Im}\,\alpha(s)   = \rho(s) \beta(s).   \label{unit} 
\ee

On the other hand, the analytical properties of the $\beta(s)$ function allow us to write it as~\cite{Chu:1969ga}
\be
\beta(s) =  \gamma(s) \hat s^{\alpha(s)} /\Gamma(\alpha(s) + 3/2) , \label{reduced} 
\ee
where $\hat s =( s-4m_\pi^2)/s_0$. The dimensional scale $s_0=1\,$ GeV$^2$ is introduced for 
convenience and the reduced residue $\gamma(s)$ is an analytic function, whose phase is known because $\beta(s)$ is real in the real axis.

Now we can write down dispersion relations for $\alpha(s)$ and $\beta(s)$, and connect them by using the unitarity condition~\eqref{unit} to obtain the following system of integral equations~\cite{Chu:1969ga}:
\begin{align}
\mbox{Re}\, \alpha(s) & =   \alpha_0 + \alpha' s +  \frac{s}{\pi} PV \int_{4m_\pi^2}^\infty ds' \frac{ \mbox{Im}\,\alpha(s')}{s' (s' -s)}, \label{iteration1}\\
\mbox{Im}\,\alpha(s)&=  \frac{ \rho(s)  b_0 \hat s^{\alpha_0 + \alpha' s} }{|\Gamma(\alpha(s) + \frac{3}{2})|}
 \exp\Bigg( - \alpha' s[1-\log(\alpha' s_0)]
+ \frac{s}{\pi} PV\!\int_{4m_\pi^2}^\infty\!\!ds' \frac{ \mbox{Im}\,\alpha(s') \log\frac{\hat s}{\hat s'} + \mbox{arg }\Gamma\left(\alpha(s')+\frac{3}{2}\right)}{s' (s' - s)} \Bigg), 
\label{iteration2}\\
 \label{betafromalpha}
 \end{align}
where $PV$ stands for ``principal value''\footnote{For the $\sigma$ meson, we make a small modification in order to include the Adler-zero required by chiral symmetry: we multiply the right hand side of Eq.\eqref{iteration2} by $2s-m_\pi^2$ (Adler zero at leading order in Chiral Perturbation Theory~\cite{chpt}) and replace the $3/2$ by $5/2$ inside the gamma functions in order not to spoil the large $s$-behavior. 
}. The phenomenological parameters $\alpha_0, \alpha'$ and $b_0$ will be determined by fitting to the resonance pole in the following way: for a given set of $\alpha_0, \alpha'$ and $b_0$ we solve the system of Eqs.~\eqref{iteration1} and \eqref{iteration2} iteratively. From the obtained Regge parameters $\alpha(s)$ and $\beta(s)$ we determine the corresponding Regge pole $\beta_M(s)/(l  - \alpha_M(s))$. The difference between the position and residue of the physical pole and the Regge one is used to define a $\chi^2$ function, which we minimize by changing the parameters $\alpha_0, \alpha'$ and $b_0$ and repeating the above steps.

\section{$\rho(770)$ and $f_0(500)$ Regge trajectories}
  
\begin{figure}
\centering
\includegraphics[scale=0.80,angle=-90]{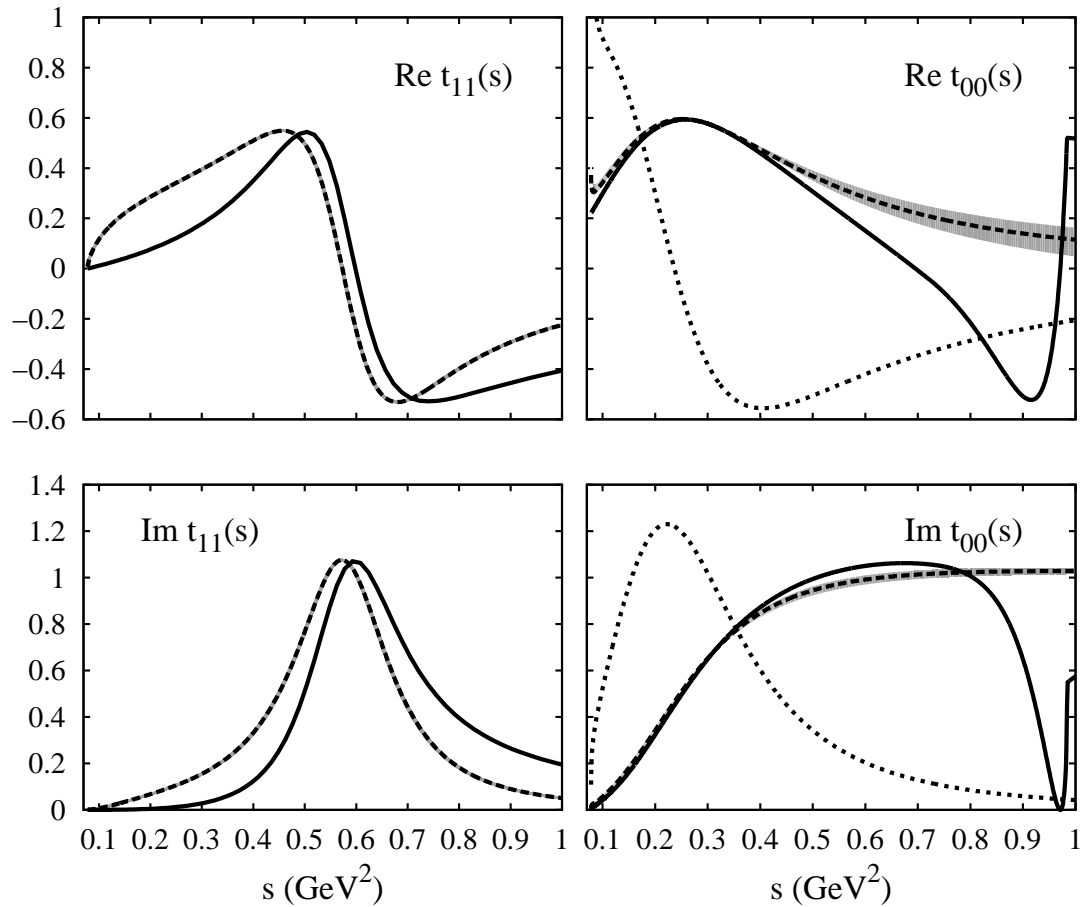}
 \caption{\rm \label{fig:ampl} 
  Partial waves $t_{lI}$ with $l=1$ (left panels) and $l=0$ (right panels). 
  Solid lines represent the amplitudes from \protect{\cite{GarciaMartin:2011jx}}, whose poles 
  are the input for the constrained Regge-pole amplitudes shown with dashed curves. The gray bands cover the uncertainties due to the errors of the inputs. In the right panels, the dotted lines represent the constrained Regge-pole amplitude for the $S$-wave 
if the $\sigma$-pole is fitted by  imposing a linear trajectory with $\alpha'\simeq 1\,$GeV$^{-2}$.}
\end{figure}

We carry out the minimization procedure explained in the previous section using as input the pole parameters from a precise dispersive representation of $\pi\pi$ scattering data \cite{GarciaMartin:2011jx}. The resulting Regge amplitudes in the real axis are shown in Fig.~\ref{fig:ampl}, where we compare them with the partial waves of \cite{GarciaMartin:2011jx}. Let us remark that they do not need to overlap since we have only constrained them at the resonance pole. However, the agreement in the resonant region is very good. As expected, it deteriorates as we approach threshold or the inelastic region, specially in the case of the $S$-wave due to the interference with the $f_0(980)$.

\begin{figure}
\centering
\hspace{-15mm}\includegraphics[scale=0.7,angle=-90]{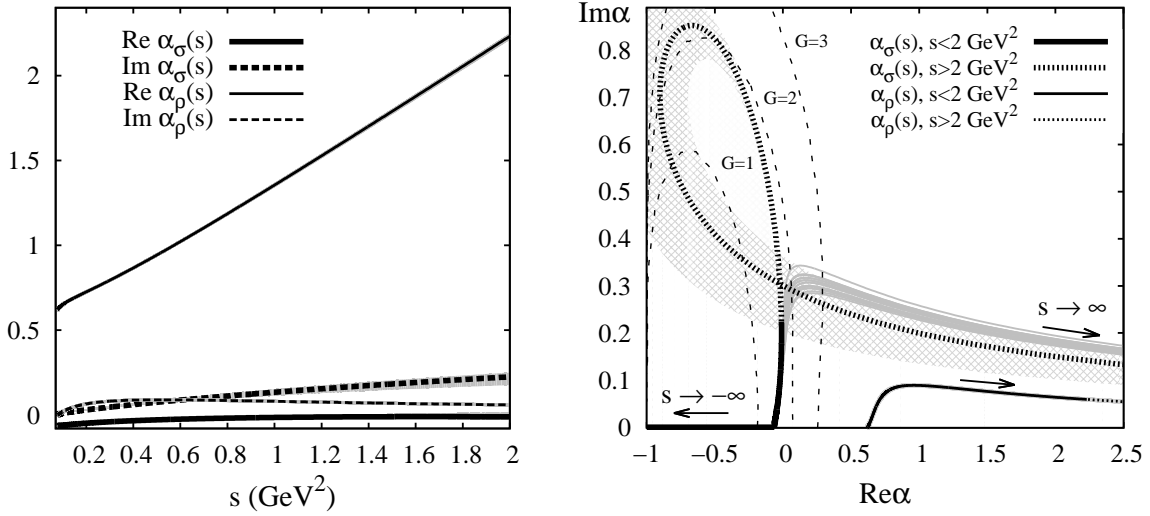}
 \caption{\rm \label{fig:trajectories-1} 
  (Left) $\alpha_\rho(s)$ and $\alpha_\sigma(s)$ Regge trajectories, 
from our constrained Regge-pole amplitudes. 
 (Right) $\alpha_\sigma(s)$ and $\alpha_\rho(s)$
in the complex plane. Beyond $s=2\,$GeV$^2$ extrapolations of our results are plotted as dotted lines. 
 Within the input pole parameter error bands, in the case of the $\sigma$, we find two types of solutions. One set (pattern-filled band)
  has a loop in the $\mbox{Im}\alpha - \mbox{Re}\alpha$ plane. The other (gray lines), having slightly higher $\alpha'$ does not form a loop. 
At low and intermediate energies, both are similar to the trajectories of the Yukawa potential $V(r)=-{\rm G} a \exp(-r/a) /r$, shown here for three different values of  G  \cite{Lovelace}. For the G=2 Yukawa curve we can estimate $a\simeq 0.5 \,$GeV$^{-1}$, following \cite{Lovelace}. This could be compared, for instance, to the S-wave $\pi\pi$ scattering length $\simeq 1.6\, $GeV$^{-1}$. }
\end{figure}

The obtained Regge parameters are given in Table \ref{aba:tbl1} and the Regge trajectories are shown in the left panel of Fig.~\ref{fig:trajectories-1}. For the $\rho(770)$ resonance we see that the imaginary part of $\alpha(s)$ is much smaller than the real part, and that the latter grows linearly with $s$. The values for the parameters are very consistent with previous determinations such as: $\alpha_\rho(0)=0.52\pm0.02$~\cite{Pelaez:2003ky}, $\alpha_\rho(0)=0.450\pm0.005$ 
\cite{PDG}, $\alpha'_\rho\simeq 0.83\,$GeV$^{-2}$ \cite{Anisovich:2000kxa}, $\alpha'_\rho=0.9\,$GeV$^{-2}$ \cite{Pelaez:2003ky}, or $\alpha'_\rho\simeq 0.87\pm0.06$ GeV$^{-2}$ \cite{Masjuan:2012gc}. This agreement is remarkable if we take into account our approximations, and that our error bands only reflect the uncertainty in the input pole parameters.

\vspace{6mm}
\begin{minipage}{\linewidth}
\centering
\captionof{table}{Parameters of the $\rho(770)$ and $f_0(500)$ Regge trajectories} \label{aba:tbl1}
\begin{tabular*}{0.7\textwidth}{@{\extracolsep{\fill} }cccc}\toprule
& $\alpha_0$ & $\alpha'$ (GeV$^{-2}$)  & \hspace{5mm}$b_0$\hspace{5mm} \\\midrule
$\rho(770)$ & $0.520\pm0.002$  & $0.902\pm0.004$ & $0.52$ \\
$f_0(500)$ &  $-0.090\,^{+\,0.004}_{-\,0.012}$ & $0.002^{+0.050}_{-0.001}$ & $0.12$ GeV$^{-2}$\\
\bottomrule
\end{tabular*}
\end{minipage}
\vspace{4mm}
  
 In the case of the $f_0(500)$ meson, we see that its trajectory is evidently nonlinear and has a slope two orders of magnitude smaller than that of the $\rho$ and other typical quark-antiquark  resonances. This strongly suggests that the nature of the $\sigma$ meson is non-ordinary. 

Furthermore, in Fig.~\ref{fig:trajectories-1} we can observe that the $f_0(500)$ trajectory is very similar to the trajectory generated by a Yukawa potential in non-relativistic scattering. Of course, our results are most reliable at low energies (thick dashed-dotted line) and the extrapolation should be interpreted cautiously. Nevertheless, our results suggest that the $f_0(500)$ looks more like a low-energy resonance of a short range potential,  {\it e.g.}\ between pions,  than a bound state of a long range confining force between a quark and an antiquark. 

In order to check that our results for the $f_0(500)$ trajectory are robust,
 we have tried to  fit the pole in~\cite{GarciaMartin:2011jx}  by fixing $\alpha'$ to a more natural value, {\it i.e.}, the one for the $\rho(770)$. With this constraint, the pole parameters are badly fitted and the resulting Regge-pole amplitude on the real axis (dotted curve in the right panel of Fig.~\ref{fig:ampl}) disagrees completely with the dispersive representation.  This shows that the large resonance width is not responsible for the fact that the $f_0(500)$ does not follow an ordinary Regge trajectory. 

\section{$f_2(1270)$ and $f_2'(1525)$ Regge trajectories}

We present here the preliminary results of our study of the trajectories of the $f_2(1270)$ and $f_2'(1525)$ resonances. In the case of the $f_2(1270)$ we use as input the pole obtained from the conformal parameterization of the D0 wave presented in~\cite{GarciaMartin:2011cn}, whereas for the $f_2'(1525)$ we use the mass and width given by the PDG~\cite{Nakamura:2010zzi}. We obtain its coupling to two mesons from that of the $f_2(1270)$ by assuming that both can be well approximated as Breit-Wigner resonances.

\begin{figure}
\begin{tabular}{cc}
\includegraphics[scale=0.75,angle=-90]{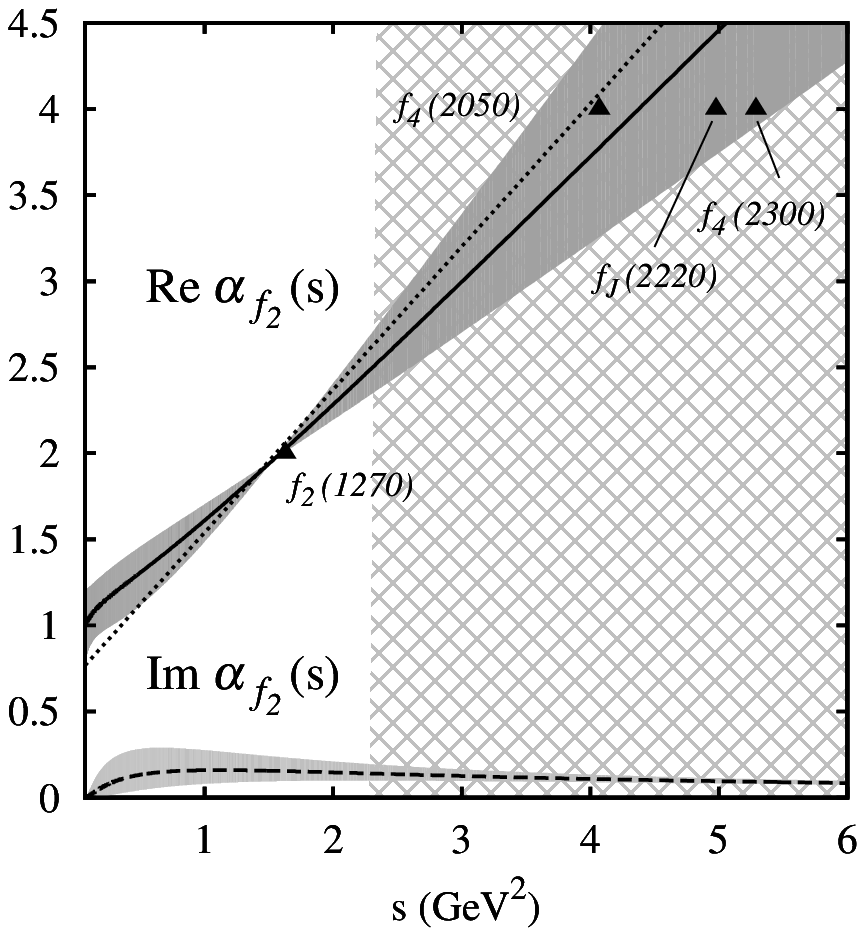}&
\hspace{-3mm}\includegraphics[scale=0.75,angle=-90]{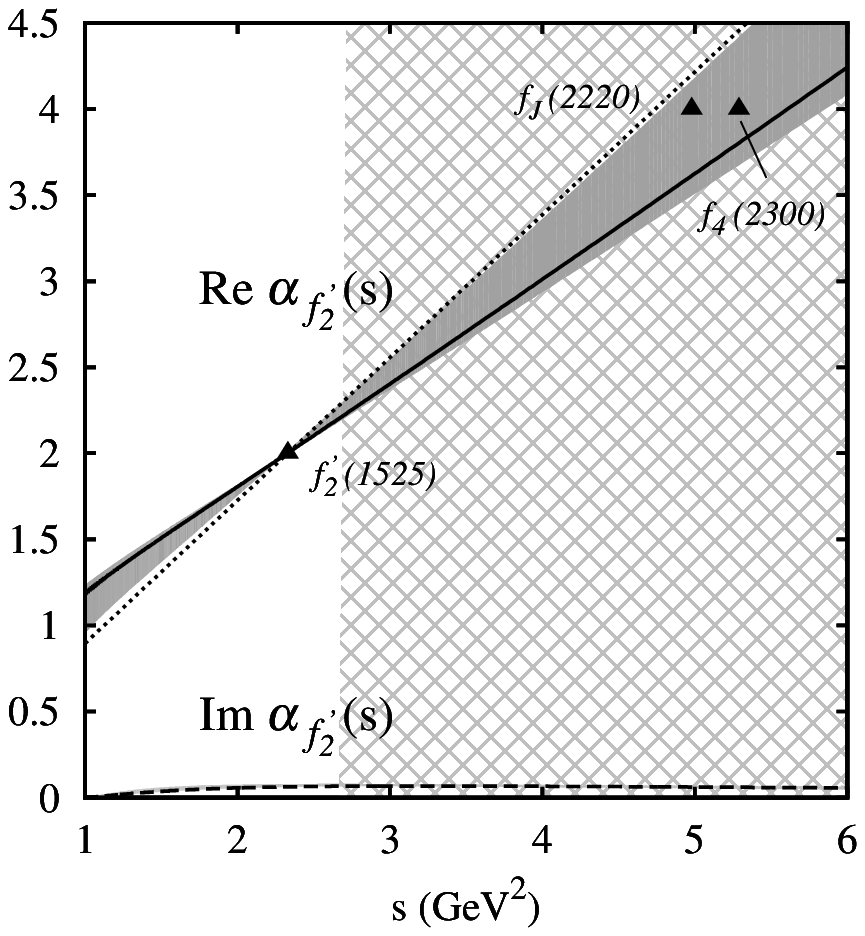}
\end{tabular}
 \caption{\rm \label{fig:trajectories-2} 
Real (solid) and imaginary (dashed) parts of the $f_2(1270)$ (left panel) and $f_2'(1525)$ (right panel) Regge trajectories.
 The gray bands cover the uncertainties due to the errors in the input pole parameters. The shaded area is the mass region three half-widths above the resonances, where our elastic approach should be considered only as a mere extrapolation. For comparison, we show with a dotted line the Regge trajectory obtained in~\cite{Anisovich:2000kxa} by fitting to meson states for leading and daughter trajectories for $P'$ (for the leading trajectory, that is to say, the one of the $f_2(1270)$, we use the intercept and slope given in the text, whereas for the $f_2'(1525)$ trajectory we take the same slope and choose the intercept for which the straight line passes by that resonance). We also show the resonances listed in the PDG that are candidates for these trajectories. Note that their average mass does not always coincide with the nominal one, as is the case for the $f_2(1270)$.}
\end{figure}

In table \ref{aba:tbl2} we show the value of the Regge parameters. They are compatible with those of~\cite{Anisovich:2000kxa}, where the authors fit all the resonances falling into the leading and daughter trajectories and obtain a common slope of $\alpha'_{P'}\approx0.83\text{ GeV}^{-2}$. Our Regge trajectories are presented in Fig.~\ref{fig:trajectories-2}, together with the trajectories found in~\cite{Anisovich:2000kxa}. We can observe that the real part of our trajectories is straight and much bigger that the imaginary part, with a slope of the order of the ``universal'' one.

In Fig.~\ref{fig:trajectories-2} we also include the resonances from the listings of the PDG~\cite{Nakamura:2010zzi} that could belong to these trajectories.  We observe that the $J=4$ resonance in the $f_2(1270)$ trajectory could be the $f_4(2050)$, as proposed in~\cite{Anisovich:2000kxa}, but also the $f_J(2220)$ or even the $f_4(2300)$. Both of these resonances appear in the particle listings of the PDG, but are still omitted from the summary tables. In fact, the former one still ``needs confirmation'' and its angular momentum is quoted to be either 2 or 4. For the $f_2'(1525)$ trajectory we find that the $J=4$ candidates are again the $f_J(2220)$ and the $f_4(2300)$. On the other hand, there is no experimental evidence of the resonance $f_4(2150)$ predicted in~\cite{Anisovich:2000kxa} from this trajectory. 

Finally, let us remark that the PDG particle listings include another $f_2$ resonance, also pending for confirmation, with a mass between that of the $f_2(1270)$ and the $f_2'(1525)$. If this resonance were to belong to a Regge trajectory with a slope similar to the values found here, both the $f_J(2220)$ and the $f_4(2300)$ resonances would be candidates for that trajectory too. For the moment we cannot apply our method to that resonance, since the branching ratios for its decay into two pions and two kaons haven't been measured yet.

\setlength\extrarowheight{5pt}
\vspace{6mm}
\begin{minipage}{\linewidth}
\centering
\captionof{table}{Parameters of the $f_2(1270)$ and $f_2'(1525)$ Regge trajectories} \label{aba:tbl2}
\begin{tabular*}{0.7\textwidth}{@{\extracolsep{\fill} }cccc}\toprule
& $\alpha_0$ & $\alpha'$ (GeV$^{-2}$)  & \hspace{5mm}$b_0$\hspace{5mm} \\\midrule
$f_2(1270)$ & $0.9^{+0.2}_{-0.3}$  & $0.7^{+0.3}_{-0.2}$ & $1.3^{+1.4}_{-0.8}$ \\
$f_2'(1525)$ &  $0.53^{+0.09}_{-0.45}$ & $0.63^{+0.20}_{-0.04}$ & $1.33^{+0.64}_{-0.07}$\\
\bottomrule
\end{tabular*}
\end{minipage}
\vspace{4mm}

\section{Summary}
We have developed a method to obtain the Regge trajectory of a resonance from its associated pole, when it dominates the elastic scattering of two hadrons in the resonance region. It is based on analytical properties of the amplitude in the complex angular momentum plane and on two coupled dispersion relations depending on some phenomenological parameters. When these parameters are fitted using as input the pole position and residue, we get a prediction for the Regge trajectory of the resonance.

We have presented here the results that we obtained in our recent work~\cite{Londergan:2013dza}, where we applied the method to the $\rho(770)$ and to the $f_0(500)$ resonances. For the $\rho(770)$ we obtain an almost real and linear trajectory, with an intercept and a slope in remarkable agreement with other values in the literature. However, for the $f_0(500)$ we get a trajectory that is highly non-linear in the resonance region, and with an imaginary part bigger that the real part. Furthermore, the real part of the trajectory has a slope more than an order of magnitude smaller than the ordinary trajectories, whereas at low energies it bears a striking similarity with the trajectories of Yukawa potentials. The resulting scale of tens or at most hundreds of MeV for the slope is more typical of meson-meson physics than of quark-antiquark interactions. These results strongly support the idea of a non-ordinary nature of the lightest scalar meson.

We have also presented our preliminary results~\cite{Prep} for the $f_2(1270)$ and $f_2'(1525)$ resonances. We obtain for both of them nearly real, straight trajectories with a slope slightly smaller than in the case of the $\rho(770)$ resonance, and compatible with a previous determination~\cite{Anisovich:2000kxa}.


\begin{theacknowledgments}
J.N. wants to thank the organizers of the conference for giving her the opportunity to present this work. J.R.P. and J.N. are supported by the Spanish project FPA2011-27853-C02-02. JN acknowledges funding by the Deutscher Akademischer Austauschdienst (DAAD), the Fundaci\'on Ram\'on Areces and the hospitality of Bonn and Indiana Universities. 
A.P.S\ is supported in part by the U.S.\ Department of Energy under Grant DE-FG0287ER40365. J.T.L. is supported by the U.S. National Science Foundation under grant 
PHY-1205019. 
\end{theacknowledgments}



\bibliographystyle{aipprocl} 

\end{document}